# High-NA In-Line Projector for EUV Lithography

Tsumoru Shintake

OIST: Okinawa Institute of Science & Technology Graduate University

This paper proposes a simple, four-mirror, in-line projector for high-NA EUV lithography that eliminates the most troublesome mask 3D effect. The design consists of a two-stage concave-convex pair, where optical aberrations are cancelled within each stage and between them, in a manner similar to that of a double-Gauss lens. The light rays pass through the central aperture in each mirror with acceptable obscuration. The numerical aperture (NA) is 0.5 and 0.7 for Hyper-NA. It has a circular exposure field with a diameter of 26 mm. The residual radial distortion is rather high at a few microns at the field rim, and the scan motion causes image blurring. Thus, we need to revert to the stepper design, and the field becomes smaller, i.e. 18 mm x 18 mm square. However, this brings an important benefit: we can remove the scanning mechanism from the photomask side. It is important to note that both the wafer and the photomask remain stationary during the EUV exposure. This guarantees superior overlay control and results in enhanced productivity. This approach serves to simplify the system and reduce electrical consumption also. Illumination will be provided through two rectangular scan-mirrors located in front of the mask, providing dual line scan field, which matches with off-axis illumination enhancing the resolution and bypasses the central obscurations.

**1. Introductions**

Non-telecentric illumination on the photomask is an EUV-specific configuration i.e. the light hits the mask at a 6-degree angle or chief-ray angle of incidence to prevent the overlap of incident and reflected light. It causes various problems, such as, mask-3D effect, can result in unwanted shadow and pattern placement shifts. In addition, there can be large differences in focus between 1D and 2D features, which limits the yield process windows.

The ideal solution to these problems is to realize telecentric illumination, just as in conventional DUV lithography. Previously, the author proposed the concept of a dual-line field, where the average normal illumination was realized and applied to two-mirror in-line EUV projector [1]. It was low-NA two-mirror projector, which covers 14 mm x 14 mm square field at 0.2 NA. It should be noted that there exists a fundamental limitation: the field size becomes quickly narrower as NA increases. This phenomenon was evident in MET3 and MET5 tools [2, 3]. The reason for this is the limited number of freedoms available to correct optical aberrations.

The EUV photolithography requires a flat field anastigmat by using only the reflection mirrors. Petzval-sum rule is the central dogma for the flat field projector. Among various candidates [2], the author focusses on the equal radii two-mirror relays as the building block, as shown in Fig. 1. They satisfy Petzval-sum rule,

$$\sum_{i=1}^{n} \frac{1}{R_i} = \frac{1}{R_1} - \frac{1}{R_2} = 0 \qquad (1)$$

Where R is the mirror curvature. By using aspheric mirrors, we may realize a flat field anastigmat with wide field.

In Fig.1, the object is located left side at center of the secondary mirror-2. The light ray starts from the object and run toward to the primary mirror-1, and bounces back to the secondary mirror-2, and finally forms an image on the primary mirror-1. By adjusting the mirror separation as listed in the figure, we may choose one of the reflection patterns. In the higher order reflections, the light ray runs outward and comes back to the center and meets on the center in high precession. We denote individual equal radii pair: 2* as two-reflection pair, 4* as four-reflection pair and 6* as six-reflection pair. There will exist even higher-order, while we have to note that EUV power will be lost in each reflection thus thy will not be feasible for EUV lithography application.

The four-reflection pair (4*) was initially studied by D. Shafer as a single-stage EUV projector in 1990 [4]. Unfortunately, a system 4 m long was required to cover a 25 mm image field, so this configuration was not studied further. The six-reflection pair was also discussed by D. Shafer in his recent presentation [5], where he suggested to utilize as the gas absorption measurement tool because of it long optical path.

To achieve higher NA with higher degree of aberration correction for a larger field size, it is essential to introduce multi-stage configuration. The image magnification factor must meet x4 in total. By simple simulation using optical design software OpTaliX [6], the author found the magnification factors x3.8, x1.9 and x1.6 for 2*, 4* and 6* reflection pairs, respectively. By looking at these factors, we found possible configuration should be a single stage 2*, two stage 4*x4* and two stage 6* x 6*. There will be another candidate: single stage 4* or 6* and two stage 2*x4*, 2*x6*, 2*/6* (second state is reversed) and 4* x 6*. By sequentially connecting them though small holes on the axis, we may build a projector with small obscuration. Among those configurations, I focused only two stage 4*x4* and two stage 6* x 6*. In other configurations, the aberration correction process on the OpTaliX did not converge well.

In this paper, all optical simulations have been performed on OpTaliX assuming perfect reflection mirrors, i.e., 100% reflectivity, no phase change associated with multi-layer coating, nor polarization dependence. The optical ray starts from the wafer (object) at left, this is opposite to the real situation. This is due to easier setup for "telecentric condition" at starting point on the wafer on OpTaliX simulator. We have to note the real system is reducer optics by factor 1/4, and the aberration value also scales as this factor.

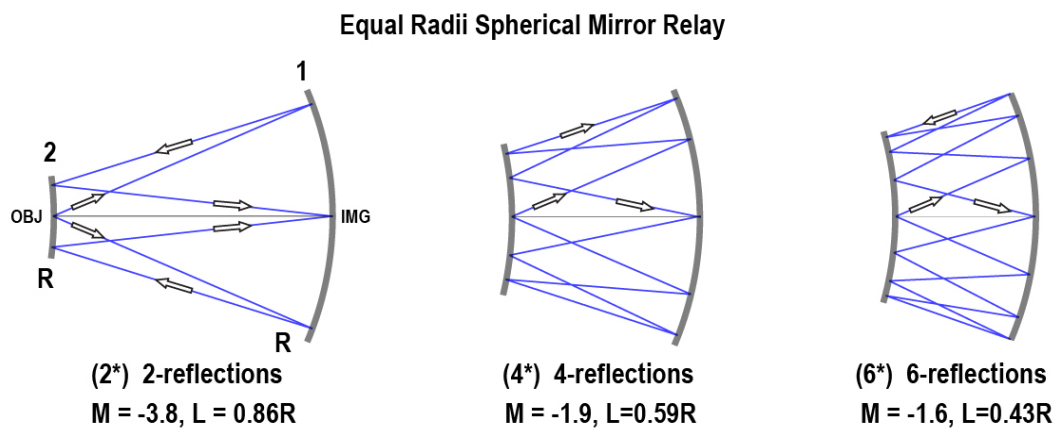

Fig. 1. Equal radii pair of two-reflection (2*), four-reflection (4*), and six-reflection (6*). We may tune the magnification factor lower or higher than the listed values by pulling the object or image vertex point outward, respectively.

## 2. Possible Configurations

Figure 2 shows possible mirror configurations for in-line EUV projector. Fig. 2 (a) is the single-stage two-reflection system: 2* proposed by the author for high efficiency EUV lithography tool [1]. It is possible to realize projector of NA 0.2 and object image distance (OID) 2000 m. It is stepper, and the field is square 14 mm x 1 4 mm. A big advantage of this system is higher transmission power efficiency, and the required EUV power becomes roughly ten times smaller than the current six-mirror EUV projection system.

Fig. 2 (b) shows two-stage configuration of four reflection pair 4* and 4* with magnification of x2.2 and x2.2 in total magnification 5, which is higher than the standard value of 4. This is due to the technical detail around intermediate image, i.e., in order to keep space between mirrors we need to pull the image point from the upstream pair to the downstream direction, resulting in increasing the magnification factor.

Fig. 2 (c) is the two-stage 6* and 6* with magnification of x2 and x2. NA 0.7 for square field size of 18 mm x 18 mm.

We have to note that there is higher EUV power loss due to multiple reflections. We need to improve reflection coefficient of the multi-layer coating and fabricate larger mirrors in high precession. Those issues may be solved by R&D on EUV technology in future.

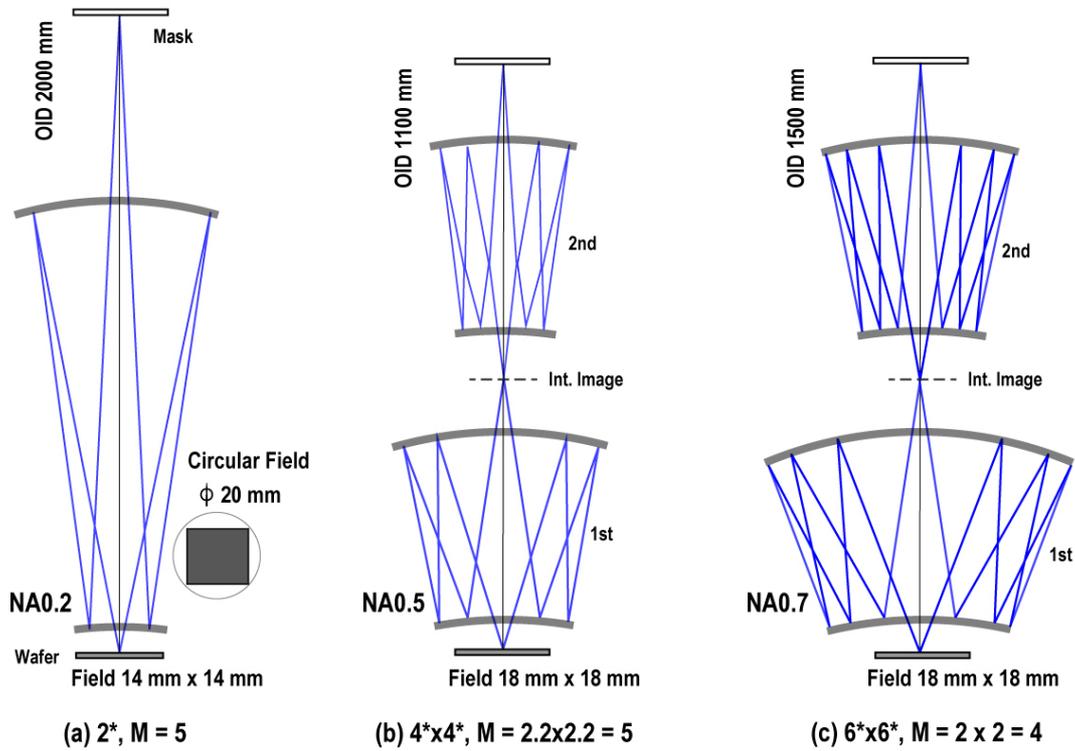

Fig. 2. Possible mirror configurations for in-line EUV projectors. (a) Single-stage, two-reflection system 2*. NA is limited to 0.2 while it has very high transmission power coefficient. To achieve wide field, the total height was designed long: OID 2000 mm. (b) Two-stage with 4* x 4* reflection. NA 0.5 with a square field 18 mmx 18 mm can be achieved. (c) Two-stage with 6* x 6* reflection. NA 0.7 with a square field 18 mmx 18 mm. Aberration correction works fairly well between 1st and 2nd stages. All optical components are axial-symmetric.

## 3. Two-stage four-reflection projector

Figure 3 shows the two-stage 4* x 4*. The numerical aperture at entrance pupil is NA 0.5, which is reduced to NA 0.23 after image magnification by factor 2.2, and thus the optical ray in the second stage becomes narrow. It crosses at intermediate image plane between two stages; thus we may design the beam hole on M4 small and minimizes the central obscuration.

As shown in Fig. 1, the four-reflection relay has magnification of -1.9, which is close to 2. In order to construct two-stage projector, we have to pull out the image vertex to downstream, which makes the magnification factor higher than 2, as a result the total magnification for two-stage cannot remain 4, it was adjusted to 5.

In this two-stage configuration, aberration from the upstream stage is cancelled in the secondary stage. Ideally, any number of reflections in upstream and downstream is possible and aberration will be corrected between two stages, similar to the double-Gauss lens. However, in practice, the computer optimization works efficiently for the same number of reflections for both stages.

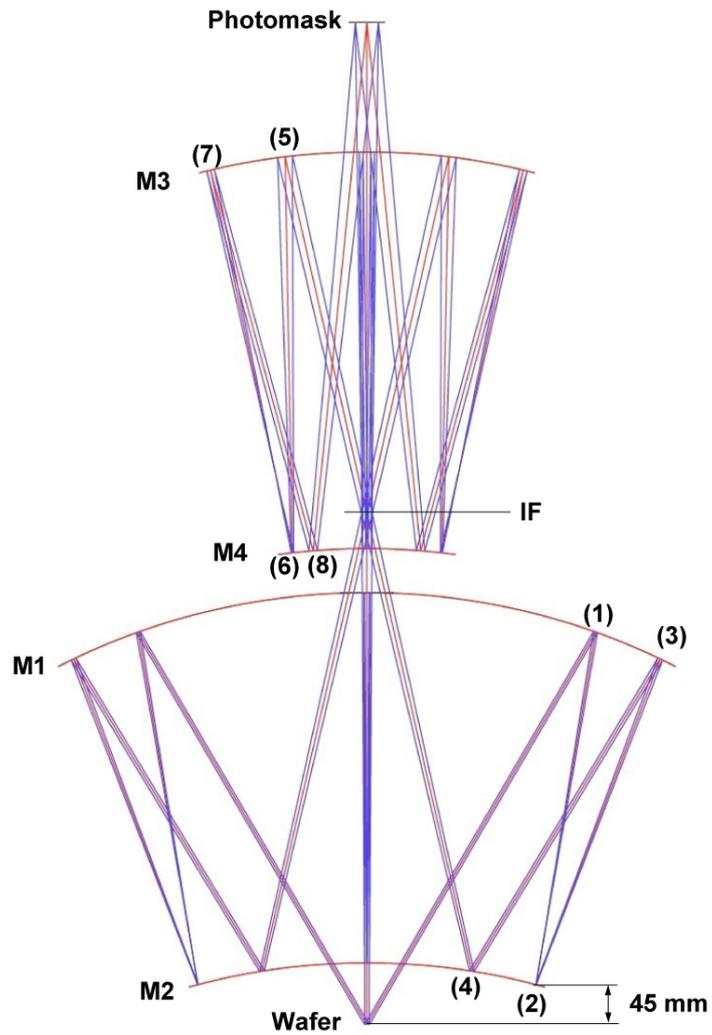

Fig. 3. Two-stage 4* x 4* projector. The optical ray crosses at the intermediate image plane between two stages near M4, thus we may design the beam hole on M4 small and minimizes the central obscuration. Note that the ray starts from the wafer in this simulation, to make the telecentric setting easier. Numbers indicated by (1) - (8) shows series of reflections.

Table-1: Parameter list for two-stage 4* x 4* projector. Note that M1~M4 have the same curvature 765 mm, which makes satisfies well Petzval condition thus we have wide field.

| NA | 0.5 | |
|---|---|---|
| Field size on the wafer | 18 x 18 | mm |
| M1 curvature, diameter | 765, 720 | mm |
| M2 curvature, diameter | 765, 400 | mm |
| M3 curvature, diameter | 765, 400 | mm |
| M4 curvature, diameter | 765, 200 | mm |
| Telecentric | wafer side | |
| Magnification | 5.0 | |
| Object image distance | 1140 | mm |
| Distortion | 0.014 | % |
| Merit | Axisymmetric Simple design | |

Table-2: Aspheric parameter.

| Mirror | A | B | C | D |
|---|---|---|---|---|
| M1 | -0.322E-10 | -0.377E-16 | -0.158E-21 | -0.154E-27 |
| M2 | -0.470E-09 | -0.222E-15 | -0.108E-19 | 0.448E-25 |
| M3 | -0.149E-09 | -0.477E-15 | 0.517E-20 | -0.286E-25 |
| M4 | -0.171E-08 | -0.118E-13 | 0.103E-17 | -0.251E-22 |

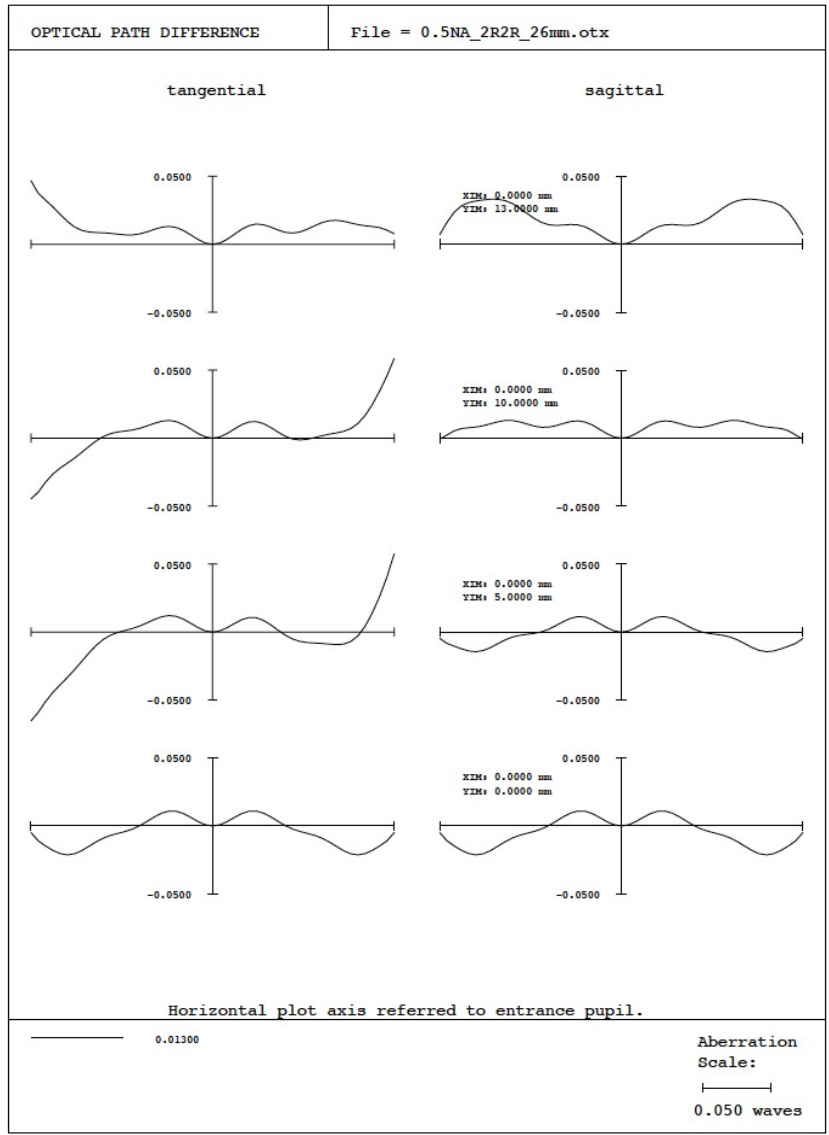

Fig. 4. Optical path difference for field points 0, 5, 10, 13 mm, they are below ±0.05 wavelength.

## 4. Two-stage six-reflection projector

This projector has ideal optical performance, i.e. it is aberration-corrected at a high level of precision up to 0.7 NA. The distortion is fairly small also, so we can use this projector as both a scanner and a stepper. Unfortunately, the optics have very high EUV power loss due to twelve reflections. Using the current Mo/Si multi-layer coating technology, the reflection coefficient is around 0.7, resulting in a total power transmission efficiency of only 0.014. In the future, when R&D on multilayer coatings has progressed, this system will become feasible.

In the current design, the mirror in the first stage is fairly large (i.e. M1, which has a diameter of 1,300 mm). This is technically infeasible and expensive. However, it is possible to scale down the first-stage optics from the wafer to the intermediate focus. While this reduces the distance from the edge of the M2 mirror to the wafer, care must be taken to avoid interruption between them.

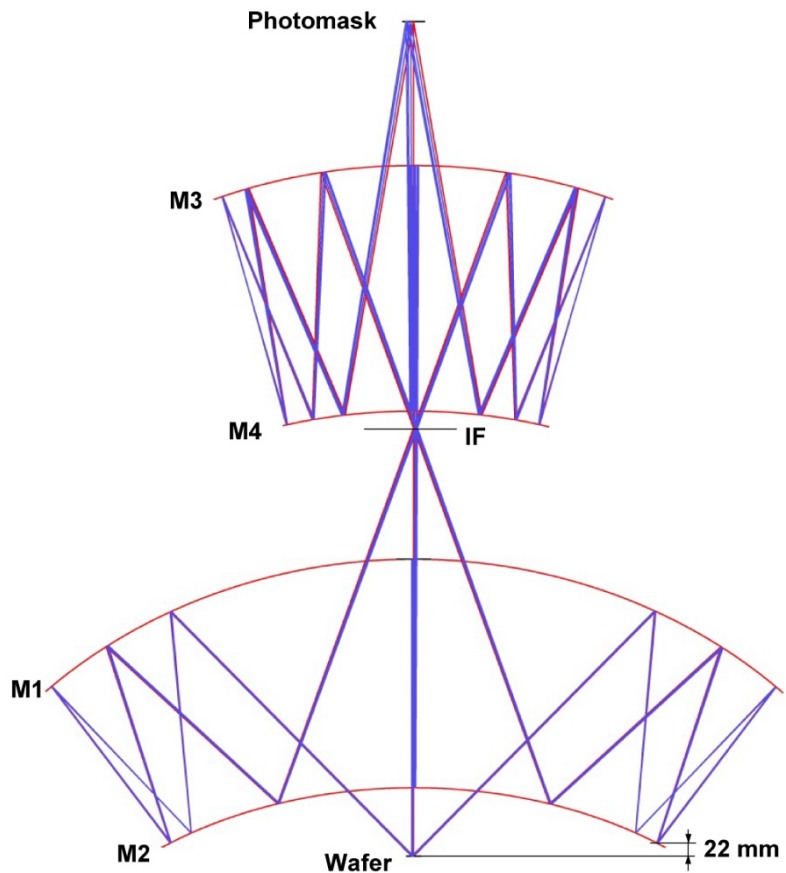

Fig. 5. Two-stage 6* x 6* projector. NA0.7, the square field 18 mm x 18 mm. OID 1500 mm. We have to note that M2 mirror edge is close to wafer level.

Table-3: Parameter list for two-stage 6* x 6* projector.

| | | |
|---|---|---|
| NA | 0.7 | |
| Field size on the wafer | 18 x 18 | mm |
| M1 curvature, diameter | 1020, 1260 | mm |
| M2 curvature, diameter | 1020, 850 | mm |
| M3 curvature, diameter | 1047, 670 | mm |
| M4 curvature, diameter | 1047, 440 | mm |
| Telecentric | wafer side | |
| Magnification | 4.1 | |
| Object image distance | 1500 | mm |
| Distortion | 0.0006 | % |
| Merit | superior aberration correction | |
| Demerit | higher EUV loss large M1 mirror higher obscuration | |

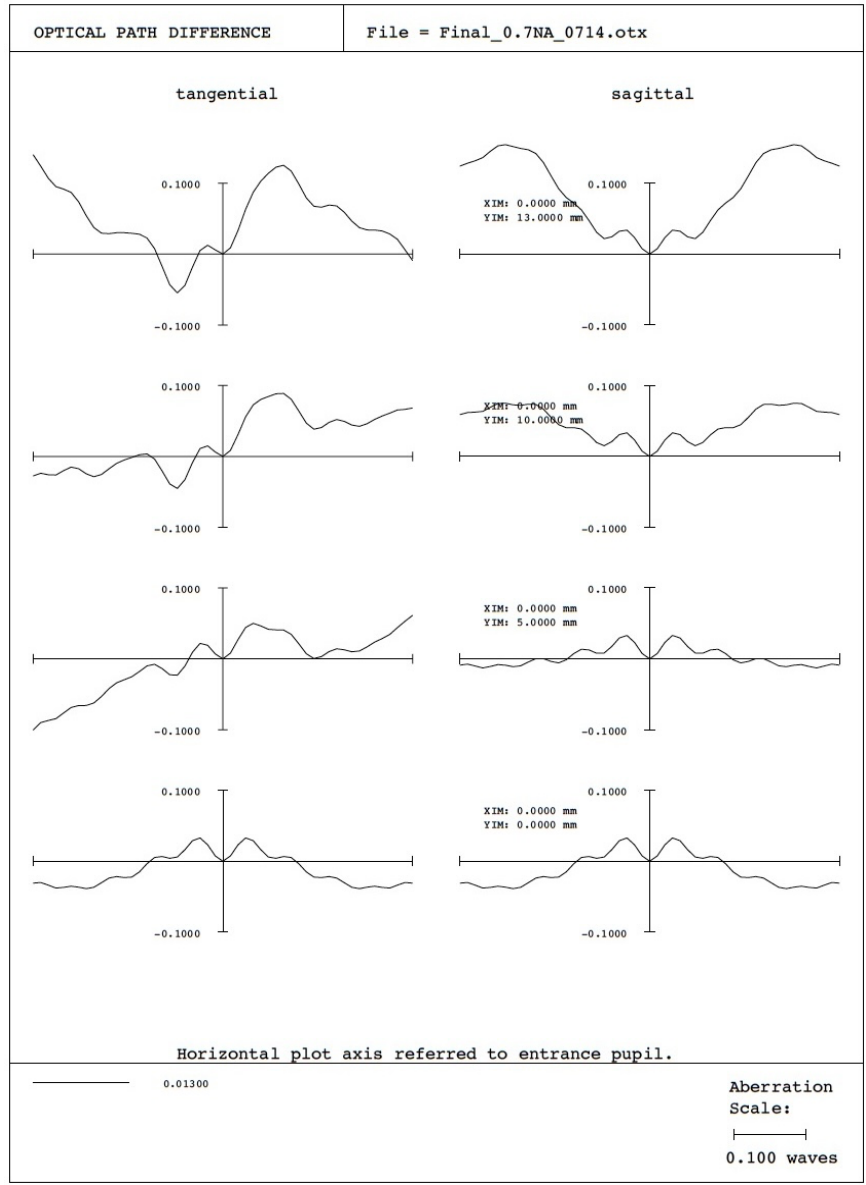

Fig. 6. Wave front error.

## 5. Central Obscuration

In an in-line projector with reflective mirrors, the illumination and diffraction from the mask pattern are carried through the central holes, which causes obscuration issues. Minimizing obscuration is therefore essential for lithography projectors. As can be seen in Figures 6 and 7, the diffraction cone bouncing off the second stage M4 mirror is the smallest size, where the central hole can cause substantial signal loss. By positioning the intermediate image close to the M4 mirror, we were able to successfully minimize the obscuration.

The remaining obscuration is primarily caused by the M2 mirror, which is positioned close to the wafer. The diffraction cone angle is large here, so a sufficiently large beam hole is required. In order to reduce the size of the beam hole, the M2 mirror surface must be brought closer to the wafer. However, to run the wafer safely, we must maintain sufficient distance from its surface. Figure 7 shows the beam hole and the diffraction cones of the second and fourth reflections on the M2 mirror. The obscuration is as low as 0.36. Further investigations will be necessary for various logic patterns and effect of partial coherency of EUV illumination. It is also possible to reduce obscuration by moving M2 surface closer to wafer (reducing gap 45 mm at the mirror edge).

Two-stage 6* x 6* projector has higher obscuration, need improvement there.

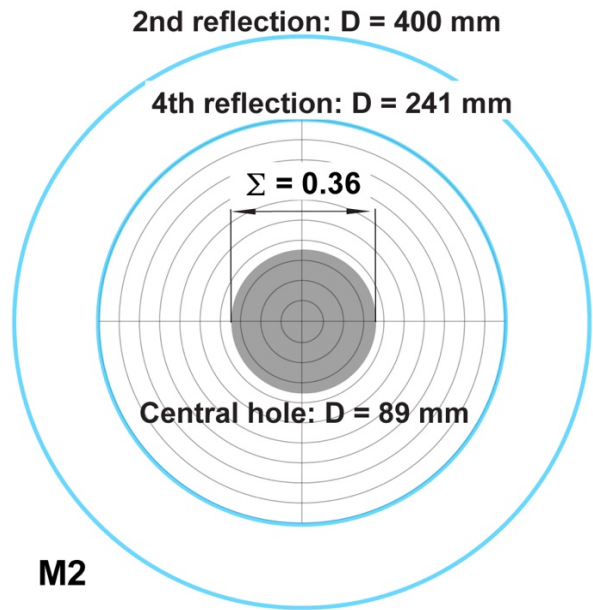

Fig. 7. Beam obscuration on M2 mirror of two-stage 4* x 4* projector. Beam holes are designed to match with the beam edge of NA 0.5, taking 1 mm gap around the beam. Obscuration is 0.36.

## 6. Illumination System

To meet the telecentric condition and eliminate the mask-3D effect, EUV light must be provided to the photomask from a normal angle. However, if an illumination mirror is placed directly in front of the photomask, the reflected light will be blocked. To solve this problem, we introduce a scan mirror, as shown in Fig. 8. This reflects the illumination light onto the photomask at tilted angles on both sides, forming a dual-line field. The reflected light from the photomask does not collide with the scan mirror and enters the projector system. By scanning the mirror along the photomask, illumination covers every point on the mask.

The dual-line field is suitable for producing dipole illumination, i.e. illumination in the form of two spots located on both sides of the NA rim at the focal plane. This provides the best resolution for vertical lines. As shown in Fig. 8(b), by mixing two components in each line field, we can form a quadrupole illumination, where four spots surround the central hole. Thus, no illumination is lost in the hole. Quadrupole illumination is widely used in conventional UV lithography due to its effective printing of various logic patterns [7].

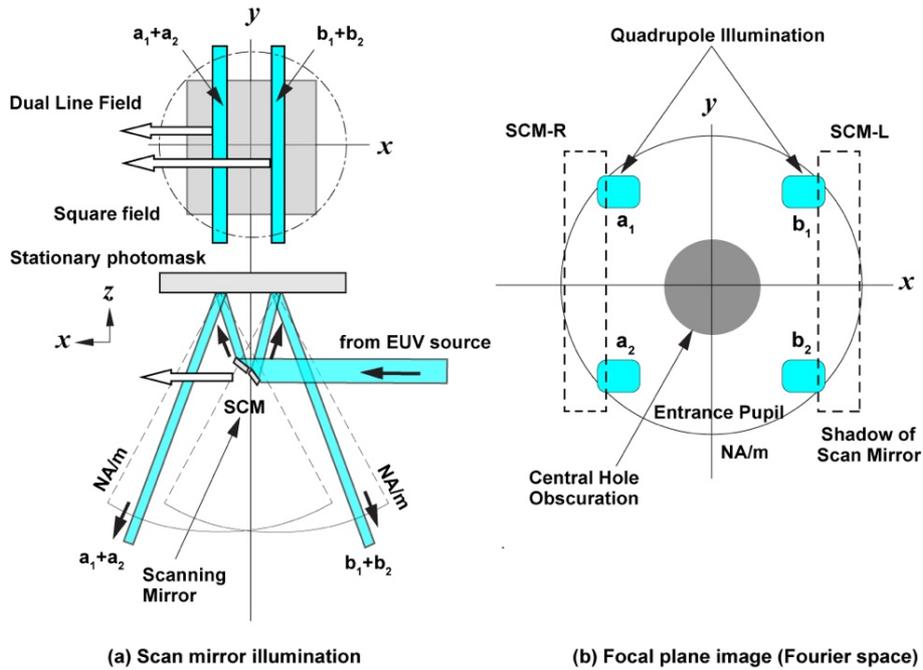

Figure 8. (a) The photomask is illuminated by a scan mirror which provides dual field lines. In each field, two components (a1+a2) and (b1+b2) are contained. (b) Four components produce quadruple pole illumination spots at the focal plane.

**7. Discussion**

The in-line projector enables the use of a stepper, meaning that the mask and wafer do not need to be scanned during exposure. Importantly, we can eliminate the expensive scanning mechanism and drive motor on the mask side. This greatly simplifies the lithography system.

In general, a scanner is preferable due to its wider exposure field. The stepper field shrinks to a square whose width is a factor of $1/\sqrt{2}$ times smaller than the scanner field width. Fortunately, the trend towards smaller chiplet designs today has resulted in die sizes of around 10 mm x 10 mm, which fit our two-stage in-line projector well.

The proposed two-stage projector requires multiple reflections. The high reflectivity of the Mo/Si layer is crucial for the success of this type of projector. Improving reflection is expected to be a focus of future R&D.